\documentstyle[prb,aps,twocolumn]{revtex}
\large
\begin{document}
\draft
\twocolumn[\hsize\textwidth\columnwidth\hsize\csname
@twocolumnfalse\endcsname

\title{High frequency dielectric and magnetic anomaly at the
phase transition in NaV$_2$O$_5$}

\author{A.~I.~Smirnov}
\address{P.~L.~Kapitza Institute for Physical Problems RAS, 117334
Moscow, Russia}
\author{M.~N.~Popova, A.~B.~Sushkov, S.~A.~Golubchik}
\address{Institute of Spectroscopy RAS, 142092 Troitsk, Russia}
\author{D.~I.~Khomskii, M.~V.~Mostovoy}
\address{Theoretical Physics and Material Science Center,
University of Groningen, 9747 AG Groningen, The Netherlands}
\author{A.~N.~Vasil'ev}
\address{M.~V.~Lomonosov Moscow State University 119899 Moscow,
Russia}
\author{M.~Isobe, Y.~Ueda}
\address{Materials Design and Charakterization Laboratory,
Institute for Solid State Physics, University of Tokyo, 7-22-1
Roppongi, Minato-ku, Tokyo 106, Japan}
\date{\today}
\maketitle

\begin{abstract}
\widetext
\leftskip 54.8pt
\rightskip 54.8pt

We found anomalies in the temperature dependence of
dielectric and  magnetic susceptibility of NaV$_2$O$_5$ in the
microwave and far infrared frequency ranges.  The anomalies occur
at the phase transition temperature $T_c$, at which the spin gap
opens.  The real parts of the dielectric constants $\epsilon_{a}$
and $\epsilon_{c}$ decrease below $T_c$.  The decrease of
$\epsilon_{a}$ (except for the narrow region close to $T_c$) is
proportional to the intensity of the x-ray reflection appearing
at $T_c$.  The dielectric constant anomaly can be explained by
the zigzag charge ordering in the $ab$-plane appearing
below $T_c$.  The anomaly of the microwave magnetic losses is
probably related to the coupling between the spin and charge
degrees of freedom in vanadium ladders.

\end{abstract}



]

\narrowtext

\section{Introduction}

Two inorganic compounds, CuGeO$_3$ and NaV$_2$O$_5$, were
extensively studied as quasi-one-dimensional (1D) magnets
containing the chains of spins $S=\frac12$  coupled by the
antiferromagnetic Heisenberg exchange.  Both materials show 1D
behavior of the susceptibility at high temperatures and the phase
transition into the spin-gap state at low temperatures.  The
lattice deformation appears simultaneously with the spin-gap
opening at the transition temperature $T_c$.  The lattice
deformation includes the doubling of the lattice period in the
spin-chain direction and the spin gap is believed to be a result
of the corresponding alternation of the exchange constants along
the chains.

The phase transition in CuGeO$_3$ is generally considered to be a
spin-Peierls transition.  The driving force of this transition is
the collective spin-lattice instability with the gain in the
energy of the spin chains exceeding the loss in the lattice
energy.\cite{Bray,Pytte} The experimental data on CuGeO$_3$ are
in a good agreement with the theoretical results for the
antiferromagnetic spin-$\frac12$ chains coupled to the
lattice.\cite{Regnault} Some deviations from the ideal
spin-Peierls behavior found for CuGeO$_3$ are also
well-understood.  They are related to a considerable interchain
interaction (the interchain exchange being about 0.1 of the
intrachain exchange\cite{Regnault}) and a strong
next-nearest-neighbor exchange (about 0.3 of the nearest-neighbor
exchange\cite{Riera}).

For some time NaV$_2$O$_5$ was considered as the second
inorganic spin-Peierls system.  The 1D-magnetic structure of
NaV$_2$O$_5$ was first associated with the chains of
V$^{4+}$-ions (spin $S$=1/2) along the b-axis of the orthorombic
crystal separated by the chains of nonmagnetic V$^{5+}$
ions.\cite{CarpyGaly,Ueda1} This structure with inequivalent
vanadium sites was recently questioned in
Refs.~\CITE{Horsch,Smolinski,Schnering,Meetsma}. On the basis of
the new structure refinement studies of the high-temperature
phase of NaV$_2$O$_5$ the new ladder-type structure with
equivalent V-sites was suggested.\cite{Horsch,Smolinski} The
ladders are oriented along the $b$-axes with the rungs along the
$a$-axis.  There is one electron per rung and the spin-$\frac12$
chains are formed by the electrons localized on V-O-V molecular
orbitals on rungs and coupled by the exchange interaction along the
$b$-direction.  In this structure the charge of V-ions
fluctuates, its average value being 4.5.

Recent experimental data,\cite{Ohama} as well as the theoretical
arguments,\cite{ThalmeierFulde,MostovoyKhomskii,Fukuyama} suggest
that the phase transition in NaV$_2$O$_5$ can be a result of some
charge ordering rather than the spin-Peierls instability.  Two
kinds of such an ordering were considered.  In the model of
Ref.~\CITE{ThalmeierFulde} the electrons below $T_c$ become
localized on one leg of each ladder, as in the initially proposed
high temperature
structure \cite{CarpyGaly}.  In this model one has to invoke an
extra mechanism to account for the spin gap at low temperatures.
In the second model \cite{MostovoyKhomskii,Fukuyama} the
electrons (or the V$^{4+}$-ions) form zigzags on each ladder.
The opening of the spin gap in this scenario is a direct
consequence of the charge ordering.  The optical spectra of
NaV$_2$O$_5$\cite{Damascelli,Popova2} were first interpreted as
an indication on the presence of inequivalent V-sites even at
room temperature.  However, these spectra could also be explained
by the presence of only a short range charge order with
relatively slow charge fluctuations.  An extra confirmation that
the phase transition in NaV$_2$O$_5$ is not an ordinary
spin-Peierls transition comes from the study of thermal
conductivity: in contrast to the spin-Peierls
system CuGeO$_3$, the thermal conductivity of NaV$_2$O$_5$ has a
huge anomaly below $T_c$, which can be naturally explained by the
charge ordering.\cite{Vasil'ev}

Further information about the charge subsystem close to the
phase transition point can be gained by studying the dielectric
constant.  Measurements of the dielectric constant at the
frequency of 1~KHz along three principal directions were
performed in Ref.~\CITE{SekineUeda}.  An anomaly with a strong
anisotropy with respect to the electric field orientation
signifying the rearrangement of the charge in NaV$_2$O$_5$ has
been found.

In this paper we present combined measurements of the real and
imaginary part of the dielectric and magnetic susceptibility of
NaV$_2$O$_5$ at the microwave frequency of 36~GHz and of the
refractive index in the far infrared range.  For comparison the
measurements of the same parameters were performed also for
CuGeO$_3$.  Our results on NaV$_2$O$_5$ confirm the presence of
strong anomalies in the dielectric constant along the $a$- and
$c$-directions, although the type of the anomaly for the
$c$-direction is different from the one found in
Ref.~\CITE{SekineUeda}.  The anomaly in $\epsilon_a$ is naturally
explained by the picture of the zigzag charge ordering suggested
in Ref.~\CITE{MostovoyKhomskii,Fukuyama}.  Possible reasons for
the anomaly in $\epsilon_c$ are discussed and further experiments
to check this picture are suggested.

\section{Experiment}

We studied the complex microwave dielectric and magnetic
susceptibilities by measuring the temperature dependence of the
resonance frequency and of the $Q$-factor of the microwave
TE$_{104}$ cavity of the size 20x7.2x3.4mm.  The resonance
frequency $f$ of the empty cavity is 36 GHz.  The presence of a
sample shifts the resonance frequency of the cavity.  When the
sample in the form of a thin plate is placed at the maximum of
the microwave electric field ${\bf E}_{mw}$, with the plane of
the plate parallel to the electric field, the resonance frequency
of the cavity should be shifted by the value
$\delta f$:
\begin{equation}
\frac{\delta f}{f}=-2 \frac{(\epsilon ^\prime -1) v}{V}.
\end{equation}
Here $v,V$ are the volumes of the sample and of the cavity
respectively, $\epsilon ^\prime $ is the real part of the dielectric
constant.

When the sample plane is oriented perpendicularly to the field
${\bf E}_{mw}$, the shift of the frequency is given by the
relation:
\begin{equation}
\frac{\delta f}{f}=-2 \frac{(\epsilon ^\prime -1) v}{\epsilon
^\prime V}.
\end{equation}
In this case the frequency shift is smaller due to the
depolarization factor effect.  Formulae (1) and (2) are valid in
the quasistatic approximation when the sample is small in
comparison with the length of the electromagnetic wave.

The imaginary part of the dielectric susceptibility $\epsilon
^{\prime \prime}$ affects the $Q$-factor and results
in a diminishing of the microwave power $U$ transmitted through
the cavity.  The relative change of $U$ is given by the relation:
\begin{equation}
\frac{\delta U}{U}=-4Q \epsilon ^{\prime \prime} \frac{v}{V}.
\end{equation}

The analogous relations could be used to determine the magnetic
permeability, $\mu=\mu ^\prime+i \mu ^{\prime \prime}$, in the
case when the sample is placed at a maximum of the microwave
magnetic field.

The sample used for the microwave susceptibility measurements had
the dimensions 1x1x0.3~mm and, thus, was much smaller than the
half of the length of the standing electromagnetic wave in the
cavity (5mm).  This fact enables us to separate the magnetic and
dielectric susceptibilities by positioning the sample at
different points of the cavity.  The $a$- and $b$-directions were
lying in the plane of the plate, the $c$-axis being perpendicular
to the plate.

The change of the refractive index $n$ with temperature was
measured in the frequency range between 1.2 and 3.0 THz by
registrating the interference pattern at different temperatures
in the transmittance spectra of thin (14-110 mkm) plates of
NaV$_2$O$_5$ single crystals cleaved perpendicular to the
$c$-direction.  The spectra were measured using the BOMEM DA3.002
Fourier transform spectrometer at the resolution of 0.2-2.0
cm$^{-1}$.  The incident light was polarized either along the
$a-$axis (${\bf E} \parallel a$) or along the $b$-axis (${\bf E}
\parallel b$).  We have also performed ${\bf E} \parallel c$
measurements at room temperature.

The position $\nu_m$ (in wavenumbers) of the m-th maximum in the
transmittance spectrum is given by the expression
\begin{equation}
\nu_m=\frac{m}{2dn},
\end{equation}
where $d$ is the thickness of the plate.  We determined the three
principal values of the refractive index along the $a$-, $b$-,
and $c$-axes by measuring the distance between the adjacent
maxima of the interference pattern.

The relative changes of the dielectric constant $\epsilon
^\prime$ were found from the measured shifts of the interference
maxima according to the relation
\begin{equation}
\frac{\delta \epsilon ^\prime}{\epsilon ^\prime}=2\frac{\delta n}{n}=
-2\frac{\delta\nu_m}{\nu_m}.
\end{equation}

 The dielectric constant is the total polarizability of the sample
divided by the sample volume.  The change of the sample size at the
phase transition  was observed by studying the thermal
expansion.\cite{Koppen} The relative  change of the  sample size in the
vicinity of the phase transition was of the order of 10$^{-4}$. In our
microwave measurements the change of the resonator frequency  is caused
by  the total polarizability of the sample.  Therefore the frequency of
the resonator is not affected by the change of the sample size if it
takes place without the change of the total polarizability.  Because we
calculate the value of $\delta \epsilon ^{\prime}$  supposing the
constant size of the sample, the experimental data  on the microwave
susceptibility  can not be attributed to the thermal expansion.

 On the contrary, the change of the refractive index measured at far
 infrared frequency by the observation of the interference pattern
 should be influenced by a change of the sample thickness and volume.
In the case of the constant total polarizability the resulting relative
change of the refractive index is to be of the order of the  relative
change of the size.

 Single crystals  of stoichiometric NaV$_2$O$_5$ used in our
experiments  have been obtained by a melt growth method
using NaVO$_3$ as a flux. The details of the growth procedure
are described in  Ref.~\CITE{IsobeKagamiUeda}. Our samples  were of
the same growth procedure as used in Ref.~\CITE{VasilevSmirnov}.  The
temperature $T_c$ of the phase transition obtained from the beginning
of the magnetic susceptibility drop  is 36$\pm 0.5$K.

\section{Experimental results}

\subsection{Microwave measurements}

The shift of the resonant frequency of the cavity produced by the
sample at T=40K was of about 400 MHz both for
${\bf E}_{mw} \parallel b$ and
${\bf E}_{mw} \parallel a$.  It corresponds to
the dielectric constant values $\epsilon^\prime _{a,b}=$12 $\pm 1$.

For ${\bf E}_{mw} \parallel c$, when the microwave electric field is
perpendicular to the plate, the shift of the resonator frequency
was only 40 $\pm$15~MHz.  Either the small value of $\epsilon
^\prime _c$ or the depolarization effect could result in this
frequency shift which is much smaller than in the case of
${\bf E}_{mw} \parallel a$ or ${\bf E}_{mw} \parallel b$.  Thus, the
absolute value of $\epsilon ^\prime _c$ could not be estimated
accurately enough: following the formula (2) we obtain the value
$\epsilon ^\prime _c$ between 4 and infinity.  However, the frequency
shift of the resonator and the value of relative change of $\epsilon_c$
at the phase transition can be measured quite accurately.  We used the
value $\epsilon ^\prime_c=$8 obtained in quasistatic
measurements\cite{SekineUeda} and in the far infrared experiments
described below, for the calculations of $\delta \epsilon _c$ according
to the formula (2).

The temperature dependence of the change of the dielectric
function was derived from the dependence of the resonator
frequency shift with respect to the frequency at T=1.5~K.  The
value of the change of the imaginary part was derived from the
temperature dependence of the transmitted signal relative to
the reference value of this signal at 40~K.

The temperature dependence of the change of the dielectric
susceptibility is shown in Fig.~1.  The real part of the
dielectric constants $\epsilon ^\prime_{a,c}$ decreases below the
phase transition temperature, while the real part of $\epsilon_b$
does not show any observable change.  The imaginary part of
$\epsilon_b$ exhibits a well-pronounced peak.  The change of the
imaginary part of $\epsilon_{a}$ and $\epsilon_{c}$ were found to
be smaller than the apparatus noise level.

A change in the response of the sample to the microwave magnetic
field was only found for the imaginary part of the magnetic
susceptibility.  These results are shown in Fig.~2.

We reexamined the CuGeO$_3$ crystals studied in
Refs.~\CITE{Smirnov1,Smirnov2} and have not found any changes of
the dielectric and magnetic susceptibilities of comparable
values.

\subsection{Far infrared measurements}

For the frequency of 1.2~THz we have obtained the following
values of $\epsilon ^\prime $ at room temperature: $\epsilon
^\prime_a$=15.0$\pm$0.6, $\epsilon ^\prime_b$=10.2$\pm$0.2,
$\epsilon ^\prime_c$=7.5$\pm$0.2.  These data are close to those
for $\epsilon ^\prime $ at zero frequency found earlier from the
fitting of the reflectivity spectra of NaV$_2$O$_5$ single
crystals by the model of independent oscillators.\cite{Popova1}

Fig.~3 shows the temperature dependence of the relative change of
the refractive index at different far infrared frequencies for
the electric field polarized along the $a$- and $b$-axes.
Simultaneously with diminishing of the refractive index $n_a$,
the absorption present in the low-frequency region of the far
infrared spectrum of NaV$_2$O$_5$ above $T_c$ decreases markedly
when temperature decreases below $T_c$ (see the inset of Fig.~3
and Ref.~\CITE{Popova2}).  We checked that despite these changes
of the absorption coefficient, equation (5) allows us to
determine the changes of $n_a$ with sufficiently good precision.
The relative change of the refractive index is of about 0.05 and
corresponds to the change of the dielectric constant that is
approximately twice as large as the value found at the microwave
frequency of 36 GHz.

We have not found any observable changes of $\epsilon^\prime_b$
(or $n_b$).  Unfortunately, we could not observe the interference
pattern for ${\bf E} \parallel c$ polarization at low
temperatures and, hence, to detect the possible changes of
$\epsilon^\prime_c$.

\section {Discussion}

The observed dielectric anomalies could, in principle, arise from
(i) an anomalous thermal expansion of NaV$_2$O$_5$ at
$T_c$,\cite{Koppen} (ii) a change of the lattice
polarizability
and (iii) a rearrangement of
electrons below $T_c$.

As it was described in Sec.II the
microwave measurements of $\epsilon$ are  insensitive to
changes of a sample size, thus the reason (i)  could
not result in the dielectric constant data shown in Fig.~1. Besides
that, the thermal expansion anomaly \cite{Koppen}  could result in
changes of $\epsilon^\prime$ of the order of 10$^{-4}$ which is 2
orders of magnitude smaller than the observed value both at microwave
and far infrared frequencies.

As for (ii), the changes in the lattice polarizability should be
accompanied by the corresponding changes in the phonon spectrum
below $T_c$.
\cite{Popova2}
  Our estimates show that the new modes that appear
below $T_c$ cannot account for the observed change of $\epsilon
^\prime$ due to their relatively small oscillator strengths.

The observed dielectric anomalies of relatively large amplitudes
agree with the conclusion about the charge ordering at $T_c$=35K
in NaV$_2$O$_5$.  Analyzing the form of the temperature
dependence of $\epsilon$ one can choose an adequate model of the
charge ordering.  The model with a localization of electrons on
one leg of the ladder\cite{ThalmeierFulde} corresponds to the
ferroelectric ordering with the spontaneous electric polarization
along the $a-$axis.  In this case the peak in the temperature
dependence of $\epsilon_a^\prime$ should be observed.  Since the
anomalies found in the real part of the dielectric constant have
the form of steps (typical for the antiferroelectric transition)
rather than peaks, we conclude that the charge ordering occurs
without a generation of a macroscopic dipole moment.  The charge
ordering of an antiferroelectric type may be in the form of the
zigzag distribution of V$^{4+}$-ions as described in
Refs.~\CITE{MostovoyKhomskii,Fukuyama}.  In this case the
decrease of the dielectric constant $\epsilon ^\prime _a$ can be
attributed to the shrinking of the electron clouds along the
$a-$direction due to the localization of electrons on the
distinct V-sites, instead of being smeared along the V-O-V
orbitals.  The observed diminishing of the optical absorption
coefficient in the frequency range studied (1.2-3.0 THz) below
$T_c$ is probably also connected with the above mentioned
shrinking of the electron clouds, the absorption itself being of
an electronic origin (see, in particular, Ref.~\CITE{Popova1}).

The coupled spin and charge degrees of freedom in vanadium
ladders were described in Ref.~\CITE{MostovoyKhomskii} using the
spin-isospin Hamiltonian.  The isospin concept is introduced for
the description of a charge ordering: Isospin $\tau_i^z=+1/2$
corresponds to the electron localized on one end of the rung $i$,
while $\tau_i^z=-1/2$ describes the electron localized on the
other end.  A charge ordering would result in the formation of
the dipole moments $d_i^z$ along $a$-direction on the rungs, with
$d_i^z \propto \tau_i^z$.  The zigzag ordering corresponds here
to $\tau^z$-antiferromagnetism.  The coupling of an external
electric field along the $a$-axis to the local (rung) order
parameter is given by the term $-E_z\tau_i^z$, similar to the
coupling $- H_z s_i^z$ in magnetic systems, and the resulting
behavior of $\epsilon ^\prime_a$ should be similar to the
behavior of the parallel magnetic susceptibility,
$\chi_{\parallel}$, in an antiferromagnet.  This consideration is
in a qualitative agreement with the experimental data shown in
Fig.~1.

Using the Landau mean field approach near $T_c$, one can show
that the decrease of the dielectric susceptibility is
proportional to the square of the charge order parameter, i.e.,
the value of $\delta \epsilon ^\prime _a$ is proportional to the
intensity of the additional x-ray reflexes that appear at $T_c$.
As shown in the inset of Fig.~1, there is, indeed, a good
correlation below $\sim(T_c-2K)$ between $\delta \epsilon _a
^\prime$ and the x-ray intensity measured in Ref.~\CITE{Fuji}.
There is some deviation of the data close to and above $T_c$ -
the value of $\delta\epsilon ^\prime_a$ does not vanish at $T_c$
and shows a tail up to $T-T_c \approx 8$K.  This deviation above
$T_c$ can naturally be ascribed to the existence of short-range
charge ordered areas which are fluctuating in some frequency
range.  These fluctuations give rise to the dynamic dielectric
susceptibility above $T_c$, but are averaged out and vanish when
the susceptibility is measured on a large time scale as it was
observed at 1KHz in Ref.~\CITE{SekineUeda}.

The nature of the behavior of the dielectric constant $\epsilon
_c ^\prime$ is less clear, especially, in view of the difference
in the anomaly of $\epsilon _c ^\prime$ obtained in our
measurements (see Fig.~1) and in the quasistatic measurements of
Ref.~\CITE{SekineUeda}, where the $\lambda-$type anomaly was
observed for $\epsilon _c ^\prime$ instead of the step-like
behavior shown in Fig.~1.

Both these measurements manifest the appearance of local dipole
moments along the $c$-direction at the transition temperature.
Possible origin of such moments may be, e.g., the shifts of the
V-ions within O$_5$-pyramides along $c$-direction.  The charge
ordering (localization of the electrons on particular V ions)
accompanied by the shifts of these ions along the $c$-direction
can modify the local dipole moments in $c$-direction.

From the crystal structure of NaV$_2$O$_5$ one would then again
expect an antiferroelectric ordering of such dipoles, which would
give the behavior of $\epsilon _c ^\prime$ consistent with the
one observed in the present work (see Fig.~1).  The reason for
different type of the $\epsilon _c ^\prime$ anomaly observed in
our measurements and in quasistatic measurements of
Ref.~\CITE{SekineUeda} is not clear at present.  A possible
explanation might be related to the small but measurable
conductivity of NaV$_2$O$_5$ single crystals which has a peak of
nearly 50 percents in amplitude at $T_c$.\cite{Loidl}
The conductivity contributes only to the
imaginary part of the microwave dielectric
constant and at 40 K the contribution is of the
order of 10$^{-7}$.  Thus, the conductivity is too small to
affect the results of the microwave measurements.  However, the
influence of this small conductivity on the results of the
electric capacity measurements at 1KHz\cite{SekineUeda} could be
significant, because the observed value of conductivity results
in the capacitor discharge time of about 10$^{-3}$ sec.
Therefore, the electric field potential within the capacitor
might be disturbed by a leakage current and this fact might
result in an additional change of the capacity.  Further
experiments are needed to elucidate this point and also to check
whether the anomalies in $\epsilon _c ^\prime $ are indeed
related to the shifts of V ions along the $c$-axis below $T_c$ as
argued above.

The picture of the phase transition in NaV$_2$O$_5$ as being
mainly of the charge-ordering type with some degree of a
short-range order persisting above $T_c$ is also qualitatively
consistent with the behavior of the dielectric losses $\epsilon
^{\prime \prime}$ shown in Fig.~1.  There should definitely exist
slow enough charge fluctuations close to $T_c$, which would give
rise to the absorption of the microwave power.  Why is the
anomaly in $\epsilon ^{\prime \prime}$ the most pronounced for
the electric field directed along $b$-axis is not clear yet.

Correspondingly, a coupling of the charge (and lattice) degrees
of freedom to spins, definitely present in this compound, should
result in magnetic losses, which were indeed observed
(see Fig.~2).  This coupling originates from the Pauli principle,
which relates the symmetry of the spatial and spin parts of the
wavefunction of two exchange-coupled electrons from the
neighboring rungs of the ladder.\cite{MostovoyKhomskii} As a
result, triplet excitations on the ladder rungs could carry a
dipole moment.\cite{Damascelli}

Note that the peak in the imaginary part of the magnetic
susceptibility is of significant value and exceeds several
times the value of the real part of the susceptibility at $T_c$.
Therefore, the fluctuations of the spin gap alone could not
explain the peak of magnetic losses and such peak is indeed not
observed in CuGeO$_3$, where the charge ordering is absent.

\section{Conclusions}

The anomalies of the dielectric and magnetic susceptibility in
the microwave and far infrared frequency range were found.  These
anomalies are in agreement with the zigzag model of the charge
ordering at the phase transition in NaV$_2$O$_5$.  The
observation of the pronounced dielectric constant anomaly and of
the peak in magnetic losses at $T_c$ indicates that the
nature of the spin gap opening in NaV$_2$O$_5$ is different from
that of CuGeO$_3$.

\section {Acknowledgments}

The work was supported by the Russian Fundamental Research
Foundation grants 98-02-16572, 98-02-17620, 96-02-19474, by
the Civilian Research and Development Foundation research project
RP1-207, by the Dutch Foundation for Fundamental Studies of
Matter (FOM), and by the OXSEN network.

{\bf Figure captions}

Fig.1 Temperature dependence of the real and imaginary parts of
the dielectric constant at the frequency 36 GHz. The inset
demonstrates
the comparison of the measurements of $\delta\epsilon _a ^\prime$
(present work) and of the intensity  $I$ of the x-ray reflexes
\cite{Fuji}:  $\alpha=1-\delta \epsilon_a ^\prime (T) / \delta \epsilon
_a ^\prime (45K)$.

Fig.2 Temperature dependence of the change of the imaginary part of
the magnetic permeability at the frequency 36 GHz.

Fig.3
Temperature dependence of the relative change of the refractive
index $n_a$ at the frequency
1.1~THz~($\Box$),
1.4~THz~($\triangle$),
1.9~THz~($\nabla$),
2.2~THz~($\Diamond$),
2.6~THz~($\circ$),
2.8~THz~($\times$),
3.1~THz~($+$) and $n_b$ at
3.1~THz~($\bullet$).
Inset presents the temperature dependence of the integrated intensity of
the far infrared absorption
$I=\int\limits_{\scriptscriptstyle1.0\rm\,THz}^{\scriptscriptstyle12
\rm\,THz} \beta(\nu)d\nu$,
$\beta$ being the absorption coefficient.

\vspace{2mm}

\end{document}